\documentstyle{article}
\title{{\bf{Nonlinear Schr\"{o}dinger Equations and  Virasoro Algebra}}}
\author{\bf{H.\ T.\"OZER}\thanks{\bf{e-mail\ :\ ozert @ itu.edu.tr}} and
\bf{S. SAL\.{I}HO\u{G}LU}\thanks{\bf{e-mail\ :\ salihogl  @ itu.edu.tr}} \\\\
 Physics Department,\ Faculty of Science and Letters,\\
Istanbul Technical University,\\
34469,\ Maslak,\ Istanbul,\\
Turkey
}
\begin{document}
\maketitle
\begin{abstract}
By using AKNS scheme and soliton connection taking values in a Virasoro algebra
we obtain new coupled Nonlinear Schr\"odinger equations.
\end{abstract}
\newpage
\setcounter{equation}{0}

\vskip 5mm

\section{\bf{Introduction}}
\par  Coupled Nonlinear Schr\"{o}dinger (NLS)
equations can be obtained using Ablowitz, Kaub, Newell, Segur (AKNS)
 scheme[1]. In this scheme in
order to obtain coupled NLS equations one way is to start by a
soliton connection which has values in sl(2,R) algebra[2]. Coupled
NLS equations on various homogeneous spaces have been obtained in
literature by assuming a soliton connection taking values in a
simple Lie algebra , in a Kac-Moody algebra and in  Lie superalgebra
[3-5].Virasoro algebra is also important in string theory, 2D
gravity,conformal field theory and KdV hierarchy[6-10]. Virasoro
algebra appears as special subalgebra of the generalized symmetry
algebra of the Kadomtsev- Petviashvili (KP) and
differential-difference Kadomtsev- Petviashvili ( D$\Delta$-KP)
equation in 2+1 dimensions [11-16].

\par In Sec.2 we will discuss the sl(2,R) algebra valued soliton connection
and we will obtain coupled NLS equations. Sec.3 concerns the
Virasoro algebra valued soliton connection and we give in this
section the new coupled NLS equations.


\setcounter{equation}{0}
\section{\bf{AKNS Scheme with sl(2,R) Algebra}}
\par  In AKNS scheme in 1+1 dimension the connection is defined as
\begin{equation}
\label{8}
\begin{array}{lll}
\Omega=& \Big(& i \lambda H_1 + Q^{+1} E_{+1}+Q^{-1} E_{-1} \Big) dx+\\
& \Big(& -A H_1 + B^{+1} E_{+1}+B^{-1} E_{-1} \Big) dt%
\end{array}
\end{equation}
\noindent where $H_1$, $E_{+1}$ and $E_{-1}$ are generators of
sl(2,R) algebra, namely they have matrix representations as

\begin{equation}
H_1 = \left[\begin{array}{cccc}
                               1 & 0\\
                               0 &-1 \end{array}
                        \right];\\
E_{+1} = \left[\begin{array}{cccc}
                               0 & 1\\
                               0 & 0 \end{array}
                        \right];\\
E_{-1} = \left[\begin{array}{cccc}
                               0 & 0\\
                               1 & 0 \end{array}
                        \right]
\end{equation}
\noindent
These generators satisfy the following commutation relations
\begin{equation}
\label{17}
\begin{array}{ccc}
 \left[ H_1,E_{\pm 1}\right] & = &\pm 2 E_{\pm 1} , \\  \left[E_{+1},E_{-1}\right] & = & H_1
\end{array}
\end{equation}
\noindent
In Eq.(1) $\lambda$ is the spectral parameter, $Q^{\pm 1}$   are fields depending on space and
time, namely x and t, and functions A,$B^{\pm 1}$ are x,t and $\lambda$ dependent.

The integrability condition is given by
\begin{equation}
\label{8}
\begin{array}{lll}
d\ \Omega\  + \ \Omega\ \wedge\ \Omega  =  0
\end{array}
\end{equation}
\par By using Eqs.(1) and (4) one can obtain following equations:
\begin{equation}
\label{18}
\begin{array}{lll}
{Q^{+ 1}}_t=\ \ {B^{+ 1}}_x+ 2 i \lambda  B^{+1} + 2 Q^{+1} A
\end{array}
\end{equation}
\noindent
\begin{equation}
\label{19}
\begin{array}{lll}
{Q^{- 1}}_t=\ \ {B^{- 1}}_x- 2 i \lambda  B^{-1} - 2 Q^{-1} A
\end{array}
\end{equation}
\noindent
\begin{equation}
\label{19}
\begin{array}{lll}
0=A_x+ B^{+ 1}Q^{- 1}-B^{-1}Q^{+ 1}
\end{array}
\end{equation}
\par In AKNS scheme we expand A,$B^{\pm 1}$ in terms of  positive powers of $\lambda$ as
\begin{equation}
A=\sum_{\scriptstyle n=0}^2 \lambda^n a_n;\ \
B^{+1}=\sum_{\scriptstyle n=0}^2 \lambda^n b^{+1}_n;\ \
B^{-1}=\sum_{\scriptstyle n=0}^2 \lambda^n b^{-1}_n
\end{equation}
\noindent Inserting Eq.(8) into Eqs.(5-7)gives 9 relations  in terms
of $a_n$,$b^{\pm 1}_n$ . By solving these relations one can get
$$a_0=- i Q^{+1}Q^{-1};\ \
a_1=0;\ \
a_2=-2 i;\ \
$$
$$
b^{+1}_0=i {Q^{+1}}_x ;\
 b^{+1}_1=2 Q^{+1};\ b^{+1}_2=0
 \eqno(9)
$$
$$
b^{-1}_0=-i {Q^{-1}}_x ;\
 b^{-1}_1=2 Q^{-1};\ b^{-1}_2=0
$$
\noindent By using the relations given by Eq.(9) from Eqs.(5-6) one
can obtain the coupled NLS equations as \setcounter{equation}{9}
\begin{equation}
\label{18}
\begin{array}{lll}
{Q^{+ 1}}_t=\ \ i \left[{Q^{+ 1}}_{xx}-2 (Q^{+ 1})^2 Q^{-1}\right]\\
{Q^{- 1}}_t=-i\left[{Q^{- 1}}_{xx}-2 (Q^{- 1})^2 Q^{+1}\right]
\end{array}
\end{equation}
\noindent If one takes $Q^{-1}=(Q^{+1})^*$ Eq.(10) becomes
\begin{equation}
\label{19}
\begin{array}{lll}
{Q^{+1}}_t= i \left[{Q^{+ 1}}_{xx}-2 (Q^{+ 1})^2(Q^{+1})^* \right]
\end{array}
\end{equation}
\noindent
Eq.(11) is called NLS equation.

\section{\bf{AKNS Scheme with Virasoro Algebra}}
\par We generalize the connection given by Eq.(1) as
\begin{equation}
\label{81}
\begin{array}{lll}
\Omega=& \Big(& i \lambda L_0 + Q^{+m} L_{+m}+Q^{-m} L_{-m}\Big)dx+\\
& \Big(& -A L_0 + B^{+m} L_{+m}+B^{-m} L_{-m}\Big)dt
\end{array}
\end{equation}
\noindent where $L_0$ and $L_{\pm m}$ are generators of centerless
Virasoro algebra, namely they satisfy the following  commutation
relations
\begin{equation}
\label{34}
\begin{array}{lll}
\left[ L_r,L_s\right]  =  (r-s)\ L_{r+s}
\end{array}
\end{equation}
\noindent In Eq.(13) r and s can be zero, or positive and negative
integers. $L_{+m}$ are generators with positive integer indices and
$L_{-m}$ are generators with negative integer indices. In Eq.(12) we
assume summation over the repeated indices. The fields $Q^{\pm m}$
are x,t dependent and functions A,$B^{\pm m}$ are x,t and $\lambda$
dependent.
 \vskip 5mm
In Virosoro  algebra if we restrict r to have only 0,+1,-1 values we
obtain sl(2,R) algebra given by Eq.(3) with following definitions:
\begin{equation}
H_1 =-2 L_0 ;\ \ E_{+1} =L_{+1} ;\ \ E_{-1} = -L_{-1}
\end{equation}
\noindent From the integrability condition given by Eq.(4) we obtain

$$
{Q^{+m}}_t= {B^{+m}}_x-i m  \lambda B^{+m}-m A Q^{m}
+\sum_{\scriptstyle r,s=1}^\infty (s-r) B^{+r} Q^{+s} \delta_{r+s,m}
$$
$$
+\sum_{\scriptstyle r,s=1\atop\scriptstyle r>s}^\infty(r+s) B^{-s} Q^{+r} \delta_{r-s,m}
+\sum_{\scriptstyle r,s=1\atop\scriptstyle r<s}^\infty(-r-s) B^{+s} Q^{-r} \delta_{-r+s,m}
\eqno(15)
$$
$$
{Q^{-m}}_t= {B^{-m}}_x+i m  \lambda B^{-m}+m A Q^{-m}
+\sum_{\scriptstyle r,s=1}^\infty (s-r) B^{-s} Q^{-r}
\delta_{-r-s,-m}
$$
$$
+\sum_{\scriptstyle r,s=1\atop\scriptstyle r<s}^\infty(r+s) B^{-s} Q^{+r} \delta_{r-s,-m}
+\sum_{\scriptstyle r,s=1\atop\scriptstyle r>s}^\infty(-r-s) B^{+s} Q^{-r} \delta_{-r+s,-m}
\eqno(16)
$$
\noindent
and
$$
0=-A_x+\sum_{\scriptstyle r,s=1}^\infty (-2 s) B^{+r} Q^{-s} \delta_{r,s}
+\sum_{\scriptstyle r,s=1}^\infty (2 s) B^{-r} Q^{+s} \delta_{r,s}
 \eqno(17)
$$
\noindent
Here $\delta$ terms are Dirac delta functions.
\vskip 3mm
\par In AKNS scheme we expand A,$B^{\pm m}$ in terms of the positive powers of $\lambda$ as
\setcounter{equation}{17}
\begin{equation}
A=\sum_{\scriptstyle n=0}^2 \lambda^n a_n;\ \
B^{+m}=\sum_{\scriptstyle n=0}^2 \lambda^n b^{+m}_n;\ \
B^{-m}=\sum_{\scriptstyle n=0}^2 \lambda^n b^{-m}_n
\end{equation}
\noindent Inserting Eq.(18) into Eqs.(15-17)gives 9 relations  in
terms of $a_n$,$b^{\pm m}_n$ (n=0,1,2). By solving these relations
we get

$$a_0=2 i \sum_{\scriptstyle r=1}^\infty Q^{+r}Q^{-r};\ \
a_1=constant=a_{10};\ \
a_2=-i;\ \
$$
$$
b^{+m}_0={-{i}\over{m}} \left[{Q^{+m}}_x-m a_{10}{Q^{+m}} \right] ;\
\ b^{+m}_1=Q^{+m};\ b^{+ m}_2=0
 \eqno(19)
$$
$$
b^{-m}_0={{i}\over{m}} \left[{Q^{-m}}_x+m a_{10}{Q^{-m}} \right] ;\
\ b^{-m}_1=Q^{-m};\ b^{-m}_2=0
$$
\noindent
By using the relations given by Eq.(19) from Eqs.(15-16) we obtain the coupled NLS equations as
$$
{Q^{+m}}_t={-{i}\over{m}}{Q^{+m}}_{xx}+i a_{10} {Q^{+m}}_{x} -2 i m
Q^{+m}\left(\sum_{\scriptstyle r=1}^\infty Q^{+r}Q^{-r}\right)
$$
$$
-i \sum_{\scriptstyle r=1\atop\scriptstyle
r<m}^\infty\left[{{2r-m}\over{m-r}}\right]Q^{+r}Q^{+(m-r)}_x
\eqno(20)
$$
$$
-i \sum_{\scriptstyle r=1\atop\scriptstyle r>m}^\infty\left[{{2r-m}\over{m-r}}\right]Q^{+r}Q^{-(r-m)}_x
+i \sum_{\scriptstyle r=1\atop\scriptstyle r>m}^\infty\left[{{2r-m}\over{r}}\right]Q^{-(r-m)}Q^{+r}_x
$$
\noindent and
$$
{Q^{-m}}_t={{i}\over{m}}{Q^{-m}}_{xx}+i a_{10} {Q^{-m}}_{x} +2 i m
Q^{-m}\left(\sum_{\scriptstyle r=1}^\infty Q^{+r}Q^{-r}\right)
$$
$$
+i \sum_{\scriptstyle r=1\atop\scriptstyle
r<m}^\infty\left[{{2r-m}\over{r-m}}\right]Q^{-r}Q^{-(m-r)}_x
\eqno(21)
$$
$$
+i \sum_{\scriptstyle r=1\atop\scriptstyle
r>m}^\infty\left[{{2r-m}\over{r-m}}\right]Q^{-r}Q^{+(r-m)}_x +i
\sum_{\scriptstyle r=1\atop\scriptstyle
r>m}^\infty\left[{{2r-m}\over{r}}\right]Q^{+(r-m)}Q^{-r}_x
$$
\noindent In Eqs.(20-21) we note that +m covers all positive
integers,-m covers all negative integers.In general the fields
$Q^{+m}$ and  $Q^{-m}$ are independent from each other. As a special case
as in Sec.2 one can choose $Q^{-m}=(Q^{+m})^*$.Since Eqs.(20-21) are derived
from the integrability condition given by Eq.(4) these coupled NLS equations
should have soliton solutions. To find these solutions for Eqs.(20-21) is an open
problem.

\section{\bf{Conclusions}}
\par The work of AKNS which is based on sl(2,R) valued soliton
connection is extended to obtain new integrable coupled Nonlinear
Schr\"odinger equations. This is achieved by assuming the soliton
connection having values in the Virasoro algebra.



\begin{thebibliography}{9}
\bibitem{1}
M.J. Ablowitz,J.D. Kaub, A.C. Newell and H. Segur, Phys. Rev. Lett, 30(1973)1262; 31(1973)125;
Stud, App. Math. 53(1974)249.
\bibitem{2}
M.Crampin, F.A.E. Pirani and D.C. Robinson, Lett. Math. Phys.2(1977)15.
\bibitem{3}
A. P. Fordy and P.P. Kulish, Commun. Math. Phys. 89(1983)427
\bibitem{4}
M. G\"{u}rses, \"{O}. O\u{g}uz and S. Saliho\u{g}lu,Int. Mod. Phys. A5(1990) 1801.
\bibitem{5}
A. Cano\u{g}lu, B. G\"{u}ldo\u{g}an, and S. Saliho\u{g}lu, Int. Mod.
Phys.A6(1991)4655; A7 (1992)7287.
\bibitem{6}
D. J.  Gross and A.A Migdal, Phys. Rev. Lett. 64(1990)127.
\bibitem{7}
M. Douglas and S. Shenker, Nucl. Phys. B335(1990)635.
\bibitem{8}
M. Douglas, Phys. Lett. B238(1990)176.
\bibitem{9}
E. Berezin and V. A. Kazakov, Phys. Lett. B236(1990)144.
\bibitem{10}
S-y Lou, Phys. Lett. B302(1993)261; A175(1993)23.
\bibitem{11}
D. David, N. Kamran, D. Levi, P. Winternitz, J. Math. Phys. 27(1986)1225.
\bibitem{12}
K. M. Tamizhmani, A. Ramani, B. Grammaticos, J. Math. Phys. 32(1991)2635.
\bibitem{13}
S-y Lou, J. Phys. A26(1993)4387.
\bibitem{14}
K. M. Tamishmani, S. Kanagavel, B. Grammaticos and A. Ramani, Chaos,
Solitons and Fractals 11(2000)1423.
\bibitem{15}
S-y Lou, C.Rogers, W. K. Schief, J. Math. Phys. 44(2003) 5869.
\bibitem{16}
X-y Tang, X-m Qian and W. Ding, Chaos, Solitons and Fractals23
(2005)1311.
\end{thebibliography}
\end{document}